\documentclass[11pt]{article} 

\usepackage[utf8]{inputenc} 

\usepackage{geometry} 
\usepackage{graphicx} 

\usepackage{booktabs} 
\usepackage{array} 
\usepackage{paralist} 
\usepackage{verbatim} 
\usepackage{subfig} 
\usepackage{amsmath} 
\usepackage{amssymb} 
\usepackage{epstopdf} 
\usepackage{color}

\usepackage{fancyhdr} 
\usepackage{sectsty}
\allsectionsfont{\sffamily\mdseries\upshape} 

\usepackage[nottoc,notlof,notlot]{tocbibind} 
\usepackage[titles,subfigure]{tocloft} 

\title{Dominating Scale-Free Networks Using Generalized Probabilistic Methods}
\author{F. Moln\'{a}r Jr.$^{1,2}$, N. Derzsy$^{1,2,}\footnote{E-mail: derzsn@rpi.edu}$, \'{E}. Czabarka$^{4,5}$, L. Sz\'{e}kely$^{4,5}$, B. K. Szymanski$^{2,3}$, G. Korniss$^{1,2}$}

\begin{document}
\maketitle

\begin{flushleft}
$^{\bf{1}}$ Department of Physics, Applied Physics, and Astronomy, Rensselaer Polytechnic Institute,
110 8$^{th}$ Street, Troy, NY, 12180-3590 USA \\
$^{\bf{2}}$ Social Cognitive Networks Academic Research Center,
Rensselaer Polytechnic Institute, 110 8$^{th}$ Street, Troy, NY, 12180-3590 USA \\
$^{\bf{3}}$ Department of Computer Science,
Rensselaer Polytechnic Institute, 110 8$^{th}$ Street, Troy, NY, 12180-3590 USA \\
$^{\bf{4}}$ Department of Mathematics,
University of South Carolina, 1523 Greene Street, Columbia, SC, 29208 USA \\
$^{\bf{5}}$ Interdisciplinary Mathematics Institute,
University of South Carolina, 1523 Greene Street, Columbia, SC, 29208 USA \\
\end{flushleft}

\section*{Abstract}
We study ensemble-based graph-theoretical methods aiming to
approximate the size of the minimum dominating set (MDS) in
scale-free networks. We analyze both analytical upper bounds of
dominating sets and numerical realizations for applications. We
propose two novel probabilistic dominating set selection strategies
that are applicable to heterogeneous networks. One of them obtains
the smallest probabilistic dominating set and also outperforms the
deterministic degree-ranked method.
We show that a degree-dependent probabilistic selection method
becomes optimal in its deterministic limit. In addition, we also
find the precise limit where selecting high-degree nodes exclusively
becomes inefficient for network domination.
We validate our results on several real-world networks, and provide
highly accurate analytical estimates for our methods.

\section*{Introduction}
It is a critical task in network science and its applications to
find methods to efficiently detect, monitor and control the behavior
of nodes in networks. Finding small dominating sets on static or
slowly evolving networks is an effective approach in achieving these
objectives. A dominating set of a network $G$ with node set $V$ is a
subset of nodes $S \subseteq V$, such that every node not in $S$ is
adjacent to at least one node in $S$, while the minimum dominating
set (MDS) is the smallest cardinality dominating set. Dominating
sets provide key solutions to various critical problems in networked
systems, such as network controllability
\cite{Cowan_2012,Nacher_SREP2013,Nacher_IEEE2013}, social influence
propagation \cite{Kelleher_1988,Wang}, optimal sensor placement for
disease outbreak detection \cite{Eubank_2004}, and finding
high-impact optimized subsets in protein interaction networks
\cite{Wuchty_PNAS2014}. The effective use of dominating sets in
these problems demands profound understanding of the behavior of
dominating sets with respect to various network features, as well as
developing effective methods for finding different types of
dominating sets that are optimal solutions for different problems.

In most applications that utilize dominating sets, the main goal is
to minimize the number of selected dominator nodes, because
implementing dominators usually incurs some form of per-node cost.
However, finding the MDS of a network is a well-known NP-hard
problem in graph theory. It was proven that finding a sublogarithmic
approximation for the size of MDS is also NP-hard, but a logarithmic
approximation can be efficiently found by a simple greedy search
algorithm \cite{Alon_1990,Alon_2000,Raz_1997}. While research is
focused on finding better approximations to the MDS
\cite{Potluri_2011,Hedar_2010} and minimum connected dominating sets
\cite{Blum_2004,Ruan_2004,Wan_2004,Chen_2009,Simonetti_2011}
(applicable to wireless communication and sensor networks), and
developing exponential algorithms to find the exact MDS
\cite{Fomin_2008,Fomin_2009,Rooij_2009,Rooij_2011}, it remains a
fundamental challenge to develop cost-efficient strategies for
selecting dominators in a network.

In this work, we consider the additional factor of local
connectivity information availability that affects the cost of
finding dominating sets. Efficient dominating set search algorithms
require full knowledge of network structure and connectivity
patterns (i.e., adjacency matrix, or equivalent adjacency
information). Obtaining this information in large networks (over
tens of millions of nodes) involves additional expenses that can
ultimately lead to overall suboptimal costs. In addition,
sophisticated search methods tend to have polynomial computational
time complexity with high orders in the number of nodes or edges,
therefore their applicability to large real networks is
questionable. Our present study is aimed towards designing
dominating set selection strategies that satisfy the cost-efficiency
demands in terms of required connectivity information, computational
complexity, and the size of the resulting dominating set. We develop
these methods for selecting dominators in heterogeneous networks,
particularly in scale-free networks, described by a power-law degree
distribution [$P(k)\sim k^{-\gamma}$]. Networks with this
fundamental property appear in numerous real-world systems,
including social, biological, infrastructural and communication
networks. Here we show that the degree-dependent probabilistic
selection method becomes optimal in its deterministic limit. In
addition, we also find the precise limit where selecting high-degree
nodes exclusively becomes inefficient for network domination.

Literature provides detailed analysis on the bounds of dominating
sets in various types of networks \cite{Haynes} with respect to
structural properties. Cooper et al. \cite{Cooper} analyzed the
behavior of MDS in model scale-free networks created by preferential
attachment rule \cite{Barabasi_1999} that generates networks with
power-law exponent of $\gamma=3$. They found that the MDS size is
bounded above and below by functions linear in $N$, where $N$
denotes the number of nodes in the network. Similar research has
been conducted on random regular graphs and Erd\H{o}s-R\'enyi (ER)
\cite{Erdos_1960} graphs. Zito \cite{Zito} studied the size of the
minimum independent dominating set on \textit{r}-regular random graphs with
$3\leq r \leq 7$ and demonstrated that the size of this set and
consequently the size of the MDS is upper bounded by a linear
function of $N$. Later, B\'ir\'o et al. \cite{Biro_2012} improved
the prefactor of the $O(N)$ bound of the size of MDS in \textit{r}-regular
graphs using a greedy algorithm
\cite{Alon_1990,Alon_2000,Arnautov_1974,Clark_1998}. In addition,
Wieland et al. \cite{Wieland_2001} derived general bounds for dense
ER graphs using fixed edge probability and demonstrated that the MDS
size scales as log $N$. However, this result cannot be applied to
sparse graphs with fixed average degrees.

Recent studies \cite{Molnar_2013,Nacher_2012} analyzed the scaling
behavior of MDS in scale-free networks with a wide range of network
sizes and degree exponents. It was found that the MDS size decreases
as $\gamma$ is lowered, and in certain special cases when the
network structure allows the presence of $O(N)$ degree hubs (when
$\gamma<2$), the MDS size shows a transition from linear to $O(1)$
scaling with respect to network size, making these heterogeneous
networks very easy to control. However, the impact of network
assortativity, which is a fundamental property in real networks, has
not been studied.

In complex networked systems, mixing patterns are usually described by
assortativity measures. A network is considered assortative if its
nodes tend to connect to other nodes which have similar number of
connections, while in a disassortative network the high degree nodes
are adjacent to low degree nodes.
Investigating the behavior of dominating sets with respect to
assortativity is essential for deeper understanding of the network
domination problem.
Several studies conducted on real-world networks have shown that social systems are assortative, while technological ones exhibit disassortative behavior \cite{Newman_2003}.
Social psychology studies have shown that humans
are more likely to establish a connection with individuals from the
same social class, or with whom they share common interests, such as
education or workplace. This tendency, named homophily, also governs
the attachment rules in real-life social systems, and it is
reflected in the mixing patterns of these networks, which are of
significant importance in dynamical processes on social networks.
Specific connectivity schemes affect influence propagation and
epidemic spread \cite{Eguiluz_2002,Eubank_2004_}, and are also
responsible for Web page ranking \cite{Fortunato_2007} and internet
protocol performance \cite{Li_2005}.

Newman proposed a method to quantify assortativity in networks using a Pearson correlation between degrees at the end of edges \cite{Newman_2003,Newman_2002}, which he
defined as the assortativity coefficient. However, a recent study of
Litvak and van der Hofstad \cite{Litvak_2013} has shown that this
coefficient has limited applicability (only for finite variances)
and is also dependent on network size. In order to resolve these
biases, they proposed a new approach to measure assortativity based
on Spearman's $\rho$ \cite{Spearman_1904}, which is a Pearson
correlation coefficient between ranked variables. This method
provides consistent assortativity values, irrespective of network
size, thus allowing assortativity comparison between various network
sizes. In addition, it can also reveal strong dependencies more
efficiently in large networks. Therefore, we also use Spearman's
$\rho$ as the assortativity measure in our work.

Here, we also develop and employ a new method to efficiently control
assortativity in network ensembles. Using this technique, our goal
is to provide a large-scale analysis on the behavior of various
dominating sets, with respect to a wide range of network parameters,
including assortativity. Finally, we also compare our findings on
model scale-free networks and real-world network samples.

\section*{Results}

We start our study by considering potential directions on how to
build dominating sets in a network without full adjacency
information. We must select nodes based solely on their individual
properties, such as the node degree, and potentially a limited
amount of global network information, such as the number of nodes,
average degree, and power-law degree exponent. We construct our
probabilistic methods (and their deterministic limit) based on this
information.

\subsection*{Probabilistic Dominating Sets}
The results of Alon and Spencer \cite{Alon_2000} provide a
graph-theoretical approach to finding an upper bound for the size of
the minimum dominating set, and as part of it they propose a
probabilistic method for selecting dominator nodes. While their
approach is theoretical, we can carry out their method, numerically,
to obtain a probabilistic dominating set, and study its properties
in scale-free networks.

\begin{figure}[tbh]
\centerline{\includegraphics[width=100mm]{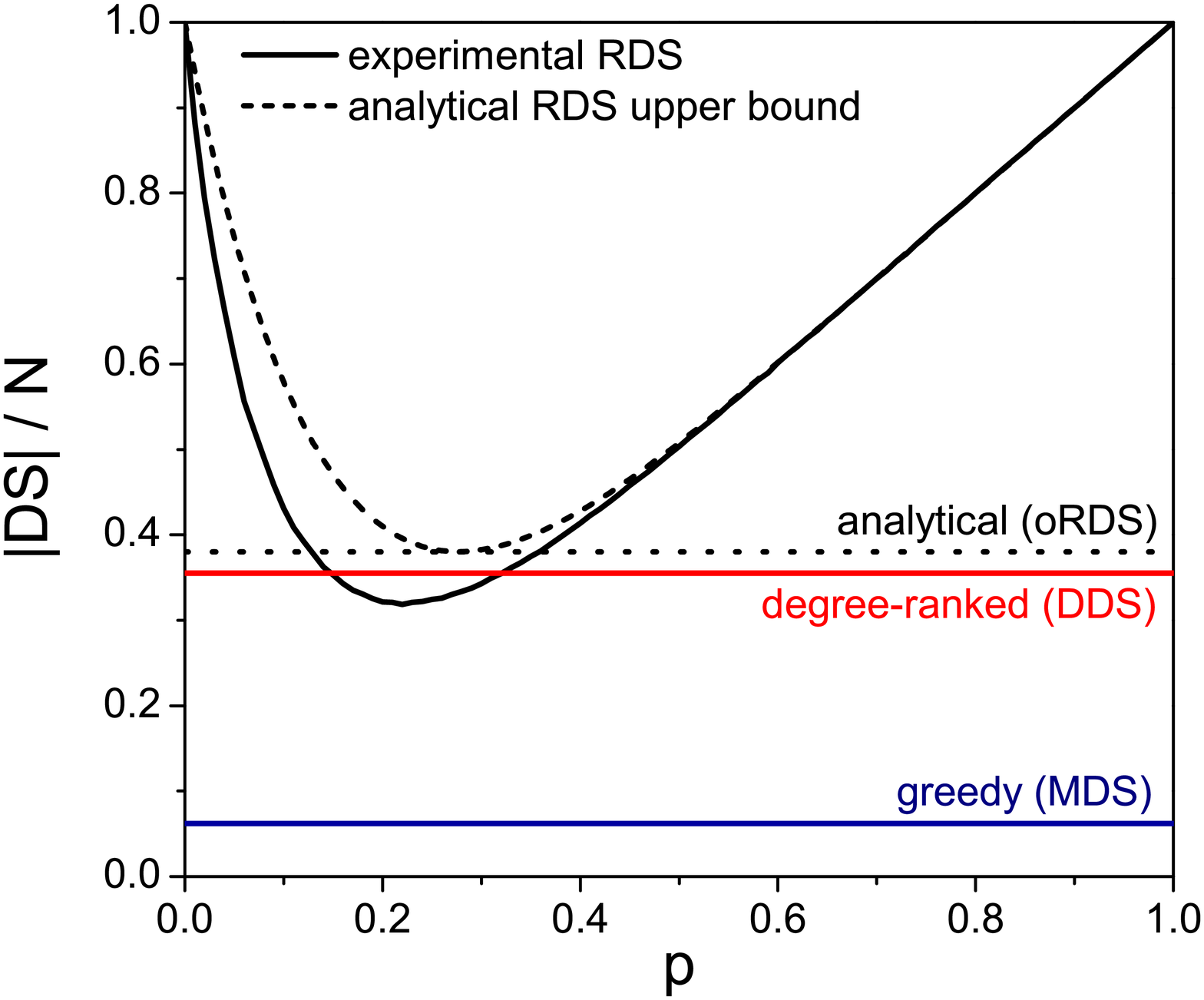}}
\caption{Analytical [black dashed curve, Eq.~(\ref{p_RDS})] and experimental [black solid curve] dominating set sizes as a function of node selection probability in random dominating sets (RDS).
The analytic optimal upper bound $|oRDS|$ [Eq.~(\ref{p_oRDS})] is indicated by the horizontal black dashed line.
The size of MDS, DDS, and the analytical estimate of the minimum of RDS is also presented for comparison. Results are averaged over 200 network realizations; $N=5000$, $\langle k \rangle=14$, $\gamma=2.5$, with $20$ dominating set searches, averaged for every network sample.}
\label{fig-upper-bound}
\end{figure}

Finding a probabilistic (random) dominating set (RDS) in a graph has
the following steps. First, we visit each node, and add it to an
initially empty set $X$, with probability $p$ (a parameter chosen
arbitrarily, $p \in [0,1]$), independently of other nodes. Then, the
remaining nodes that are not in $X$ nor adjacent to any node in $X$
are placed in set $Y$. The dominating set is obtained by $X \cup Y$.
Alon and Spencer showed \cite{Alon_2000} that the expected size of
this set is
\begin{equation}
|RDS| = |X|+|Y| \leq Np+N(1-p)^{k_{\min}+1} \;,
\label{p_RDS}
\end{equation}
where $k_{\min}$ is the minimum degree and $N$ is the number of
nodes in the graph. By differentiation of this formula with respect
to $p$ we can find the optimal $p$ value that minimizes $|RDS|$ (the
corresponding dominating set is denoted by oRDS; o stands for
optimal), which is then further bounded from above:
\begin{equation}
|MDS| \leq |oRDS| \leq N[1-k_{\min}(1+k_{\min})^{-1-1/k_{\min}}] \;.
\label{p_oRDS}
\end{equation}
Our numerical results on scale-free network samples in comparison with the analytical values is shown in Fig.~\ref{fig-upper-bound} for one set of network parameters, while plots over a wide range of parameters ($2 \leq \gamma \leq 4$; $4 \leq \langle k \rangle \leq 16$) are included in the Supplementary Information, Figures S1 and S2. We find that our numerically obtained RDS size is substantially lower than the analytical one, for most studied combination of network parameters. However, when $p\stackrel{>}{\sim}0.5$ the size of the RDS found numerically closely approaches the analytical curve, in every case.

We also compare the size of RDS to other dominating sets. We find
that the sequential greedy method that approximates the MDS (and
uses adjacency information) is still far more efficient than RDS.
However, we find that the simple degree-ranked dominating set (DDS)
is outperformed by RDS for optimally chosen $p$ values.

\subsection*{Degree-Dependent Random Dominating Sets}
To improve the results of RDS we have to consider that complex
networks are heterogeneous, and it would be beneficial to exploit
this characteristic in the probabilistic node selection strategy. We
propose a novel degree-dependent probability function for selecting
nodes that are placed in set $X$:
\begin{equation}
p_i = \min\left\{ 1,p \left(\frac{k_i}{k_{\max}} \right)^{\beta} \right\} \;,
\label{p_min1}
\end{equation}
where $k_i$ is the degree of node $i$, $k_{\max}$ is the maximum
degree in the network, and $p$ and $\beta$ are parameters. Note that
we no longer require $p$ to be a probability but rather a prefactor
that can have any positive value. Similarly to the case of
degree-independent selection probability, set $Y$ contains nodes
that are not dominated by $X$, and the ultimate result, RDS is
obtained by $X \cup Y$. The main feature of the new node selection
probability is that nodes with higher degrees are more likely
selected, which is generally desired to lower the total number of
dominators. Note, that when $p>1$, we can have $p_{i}=1$, in which
case node $i$ is surely selected.

\begin{figure}[h!]
\centerline{\includegraphics[width=100mm]{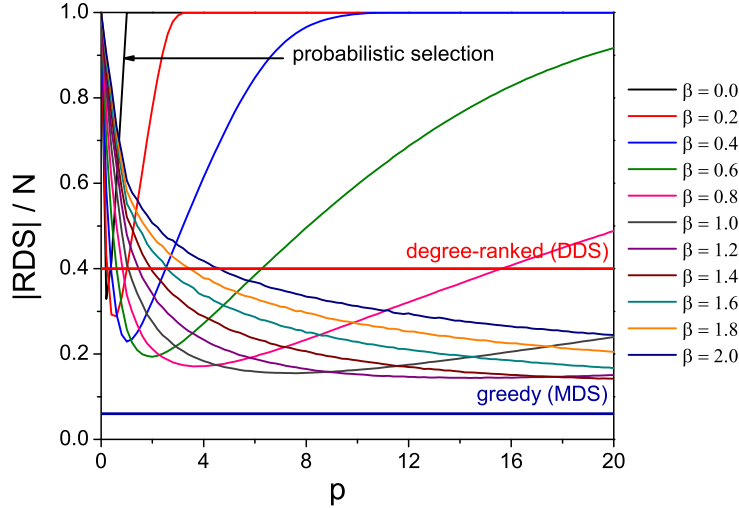}}
\caption{Size of random dominating sets (RDS) as a function of $p$ prefactor in the degree-dependent node selection probability [Eq.~(\ref{p_min1})]. Data is averaged over 200 network samples and 20 repetitions of dominating set searches for each sample. Network parameters: $N=5000$, $\langle k \rangle =14$ and $\gamma=2.5$.}
\label{fig-RDS}
\end{figure}

Figure \ref{fig-RDS} compares RDS with degree-dependent and
degree-independent node selections for a wide range of $\beta$
values (note, $\beta=0$ is identical to the degree-independent
case). In agreement with our expectations, our results clearly show
that degree-dependent node selection provides a much smaller
dominating set than the simple degree-independent selection, and
thus it also outperforms the degree-ranked selection. We can also
observe that as the $\beta$ parameter is increased the smallest
possible RDS size decreases, and it approaches the greedily
approximated MDS size. Notice however, that for finding the the
smallest possible RDS the value of $p$ has to increase as well.

\subsection*{Cutoff Dominating Sets (CDS)}

Since the smallest RDS size obtained seems to become lower for ever
increasing $\beta$ values, we expect to find the minimum with $\beta
\rightarrow \infty$. Notice, that in this case all nodes with degree
$k_i > k_{\max}p^{-1/ \beta}$ are selected with probability $1$ and
nodes with smaller degrees are selected with probability $0$. Thus,
we have a degree threshold, $\kappa \equiv k_{\max}p^{-1/ \beta}$
that now deterministically decides whether nodes will be added to
set $X$ or not. Notice, that we can use this $\kappa$ to
reparametrize the node selection probability in RDS as well:
Eq.~(\ref{p_min1}) now becomes
\begin{equation}
p_i = \min\left\{ 1,\left( \frac{k}{\kappa} \right)^{\beta} \right\} \;.
\label{p_min2}
\end{equation}
This form shows even more explicitly that the $\beta \rightarrow
\infty$ case transforms the probabilistic selection into a
deterministic one based on the $\kappa$ degree cutoff. Therefore, we
call the final result a cutoff dominating set (CDS).

\begin{figure}[h!]
\centerline{\includegraphics[width=100mm]{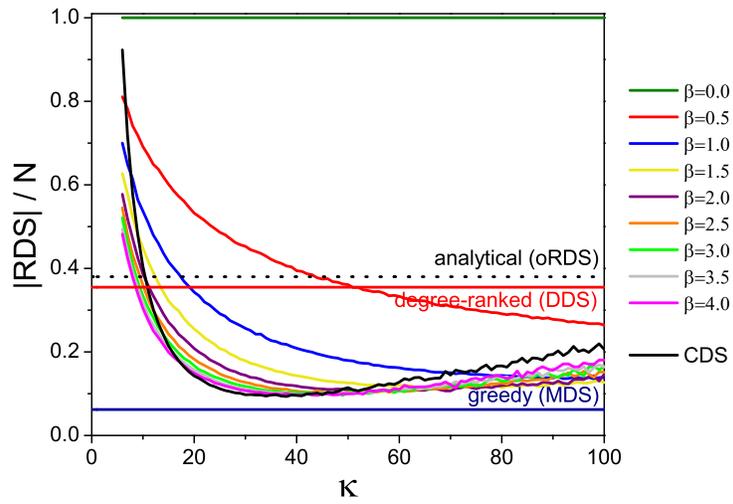}}
\caption{Cutoff dominating set (CDS) as a function of $\kappa$ degree cutoff parameter in the degree-dependent node selection probability [Eq.~(\ref{p_min2})]. For comparison, curves of RDS are plotted for various $\beta$ values. CDS corresponds to $\beta=\infty$.}
\label{fig-CDS}
\end{figure}

Figure \ref{fig-CDS} shows CDS in comparison with RDS for various $\beta$ values; similar plots for a wide range of parameters ($2 \leq \gamma \leq 4$; $4 \leq \langle k \rangle \leq 16$) are included in the Supplementary Information, Figures S3 and S4. We can see that CDS indeed provides the smallest dominating set size among probabilistic methods, and when $\kappa$ is optimal (i.e., it minimizes the size of CDS) the size of CDS almost reaches the greedy MDS approximation. Considering how much simpler CDS is compared to the greedy approximation, this result is quite remarkable.

\begin{figure}[h!]
\centerline{\includegraphics[width=\textwidth]{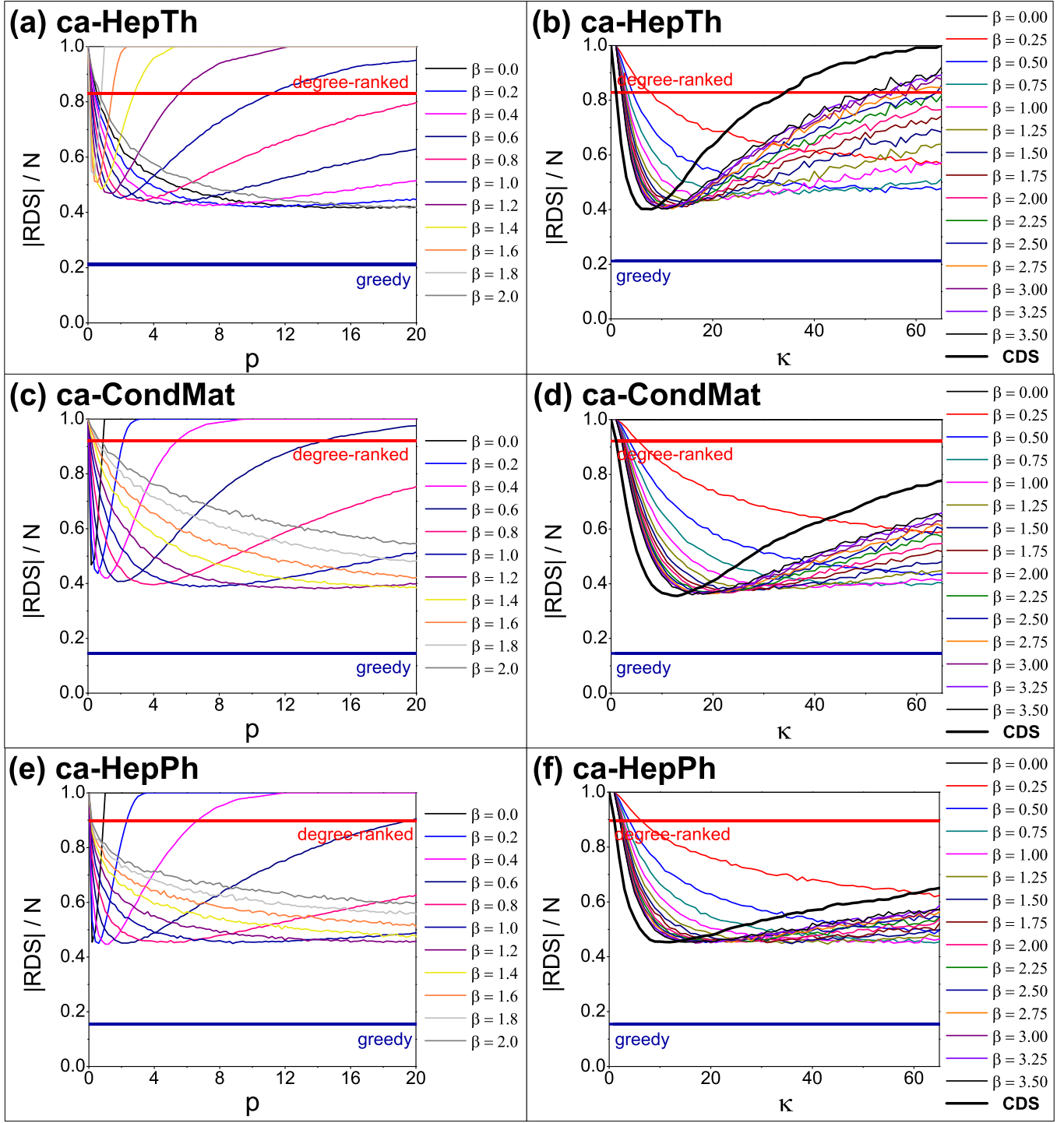}}
\caption{Sizes of probabilistic dominating sets (RDS) in real networks. Subfigures (a), (c) and (e) show the RDS size as a function of $p$ prefactor of node selection probability [Eq.~(\ref{p_min1})], while subfigures (b), (d) and (f) show the same as function of $\kappa$ degree cutoff [Eq.~(\ref{p_min2})]. For comparison, the degree-independent probabilistic ($\beta=0.00$), degree-ranked and greedy dominating set sizes are also plotted. Network parameters: ca-HepTh: $N=8638$, $\langle k \rangle=6$, $\gamma=2.2$; ca-CondMat: $N=21363$, $\langle k \rangle=8$, $\gamma=2.7$; ca-HepPh: $N=11204$, $\langle k \rangle=21$, $\gamma=1.7$.}
\label{fig-real}
\end{figure}

In order to further validate the performance of CDS, we calculate it on several real-world network samples and compare it to RDS, as well as greedy MDS approximation and DDS. We use three collaboration networks from the Stanford large network dataset collection \cite{Stanford}, namely the scientific co-authorship networks of Astro Physics (ca-AstroPh), Condense Matters (ca-CondMat) and High Energy Physics (ca-HepPh). Figure \ref{fig-real} shows these results. In all cases, we see the same behavior of CDS as in synthetic networks: CDS reaches the smallest possible size of all probabilistic dominating sets, and in some cases, it can get very close to the greedy MDS approximation.

\subsection*{Analytical Estimates of RDS and CDS}
Since both RDS and CDS require only the degree of each node to decide whether to place that node in the $X$ set, we can estimate the size of RDS and CDS in the infinite network size limit using continuous degree distributions. In general, we can estimate the size of any probabilistic dominating set in a network with any degree distribution and degree correlations as follows:
\begin{equation}
\frac{\langle DS \rangle}{N}=\int_{k_{\min}}^{k_{\max}}X(k)P(k)\mathrm{d}k+\int_{k_{\min}}^{k_{\max}}(1-X(k))P(k)\left[\int_{k_{\min}}^{k_{\max}}(1-X(k'))P(k'|k)\mathrm{d}k'\right]^{k} \mathrm{d}k,
\label{base}
\end{equation}
where $P(k)$ is the degree distribution on the domain of $[k_{\min}, k_{\max}]$, $X(k)$ is the probability of selecting a node with degree $k$ into set $X$, $P(k'|k)$ is the degree distribution of the neighbors of a node with degree $k$. The first integral calculates the expectation of $|X|/N$, while the rest is the expectation of $|Y|/N$. The latter is obtained by counting the nodes that are not in $X$ (the first part), but only those that also have no neighbors in $X$ (the expression in square brackets).

We can plug in the properly normalized power-law degree distribution in $P(k)$. Further, for uncorrelated networks we have $P(k'|k)= k' P(k') / \langle k \rangle$. For RDS with uniform node selection probability we have $X(k)=p$, resulting in:
\begin{equation}
\frac{\langle RDS \rangle}{N}=p+\frac{(1-p)(1-\gamma)}{k_{\max}^{1-\gamma}-k_{\min}^{1-\gamma}}[k_{\min}^{1-\gamma}E_{\gamma}(-k_{\min} \log (1-p))-k_{\max}^{1-\gamma}E_{\gamma}(-k_{\max} \log(1-p))]
\label{RDS}
\end{equation}
For RDS with degree-dependent probability we have $X(k)=\min(1,(k/\kappa)^{\beta})$, resulting:
\begin{equation}
\frac{\langle RDS \rangle}{N} = \frac{k_{\max}^{1-\gamma} - \kappa^{1-\gamma} + (1-\gamma) \left[y_{1} + \kappa^{-\beta}(x + y_{2}) \right]}{k_{\max}^{1-\gamma}-k_{\min}^{1-\gamma}},
\label{RDS_degree}
\end{equation}
with
\begin{eqnarray}
x & = & \frac{\kappa^{1 + \beta - \gamma}-k_{\min}^{1 + \beta - \gamma}}{1+\beta-\gamma} \\
y_1 & = & k_{\min}^{1-\gamma}E_{\gamma}(-k_{\min}\log a) - \kappa^{1-\gamma}E_{\gamma}(-\kappa \log a) \\
y_2 & = & \kappa^{1+\beta - \gamma}E_{\gamma - \beta}(-\kappa \log a) - k_{\min}^{1 + \beta - \gamma}E_{\gamma - \beta}(-k_{\min} \log a) \\
a & = &\frac{\kappa^{2-\gamma}-k_{\min}^{2-\gamma} }{k_{\max}^{2-\gamma}-k_{\min}^{2-\gamma} } - \frac{(2-\gamma)\kappa^{-\beta}}{2+\beta-\gamma} \left( \frac{\kappa^{2+\beta-\gamma}-k_{\min}^{2+\beta-\gamma}}{k_{\max}^{2-\gamma}-k_{\min}^{2-\gamma}} \right).
\end{eqnarray}
Finally, for CDS we have $X(k)= \Theta(k-\kappa)$, where $\Theta$ is the Heaviside step function that returns $1$ for positive arguments and $0$ otherwise, yielding:
\begin{equation}
\frac{\langle CDS \rangle}{N}=\frac{k_{\max}^{1-\gamma}-\kappa^{1-\gamma}+(1-\gamma)[k_{\min}^{1-\gamma}E_{\gamma}(-k_{\min}\log b)-\kappa^{1-\gamma}E_{\gamma}(-\kappa\log b)]}{k_{\max}^{1-\gamma}-k_{\min}^{1-\gamma}},
\label{CDS}
\end{equation}
with
\begin{equation}
b = \frac{\kappa^{2-\gamma}-k_{\min}^{2-\gamma}}{k_{\max}^{2-\gamma}-k_{\min}^{2-\gamma}}.
\end{equation}
Note, that in all the above formulas, $E_{n}(z)$ denotes the exponential integral function, $E_{n}(z) = \int_{1}^{\infty} e^{-zt} t^{-n} \mathrm{d}t$. The detailed derivation of the analytical estimates can be found in Supplementary Information, Section S.3.

Figure~\ref{fig-analytic-real} shows the accuracy of our analytical estimates in comparison with the numerical results of RDS and CDS. Further results on scale-free networks with different $\langle k \rangle$ and $\gamma$ values are provided in the Supplementary Information, Section S.3.4, showing that as the $\langle k \rangle$ increases, the accuracy of the analytical estimates improves. For CDS and degree-independent RDS the estimates are very close to the numerically obtained values, even with a small $\langle k \rangle$. The estimates for degree-dependent RDS are slightly less accurate, but still sufficient to provide a useful approximation of the expected dominating set size. Therefore, we can easily calculate a very accurate expected size of these dominating sets in uncorrelated scale-free networks, based on nothing beyond basic network parameters.

\subsection*{Effects of network assortativity}

\begin{figure}[h!]
\centerline{\includegraphics[width=\textwidth]{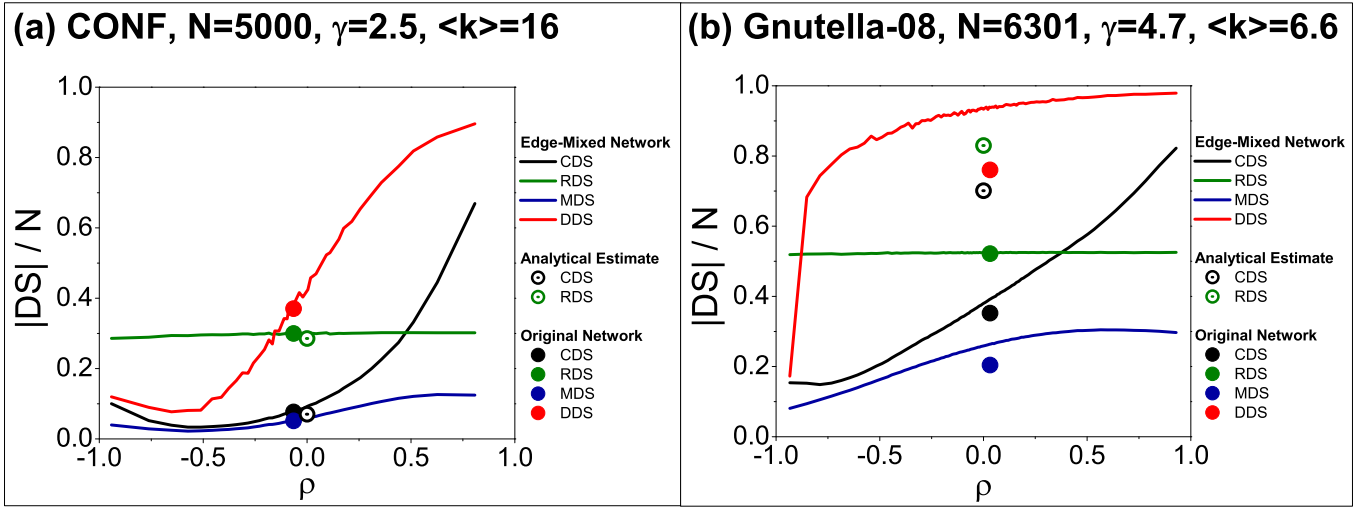}}
\caption{Dominating sets as a function of Spearman's $\rho$ assortativity measure in
(a) a synthetic network and (b) a real-world network (Gnutella).
Networks with assortativity values different from the original network are obtained by guided edge-mixing
with double-edge swaps.}
\label{fig-ds-rho}
\end{figure}

Using our edge-mixing method to control the assortativity of a network, we have compared the sizes of dominating sets as a function of assortativity, measured by Spearman's $\rho$. Figure~\ref{fig-ds-rho} shows our results for a synthetic network and a real social network, while the same comparison for different network parameters is provided in the Supplementary Information, Section S.5 for artificial networks, and Section S.6 for real networks.

As expected, the size of most dominating sets increase with higher assortativity, except for RDS with degree-independent selection probability. The most dramatic size increase is observed in DDS, which indicates that this method can only be considered viable in real-world applications for highly disassortative networks. Also, as the assortativity increases, CDS becomes larger than the simple RDS at a certain point, indicating that favoring high-degree nodes as dominators is not an effective strategy when the network is highly assortative. While the MDS size obtained by greedy search also increases with increasing assortativity, it shows the smallest increase, thus the advantage of greedy search over other methods is more pronounced.

\begin{figure}[h!]
\centerline{\includegraphics[width=140mm]{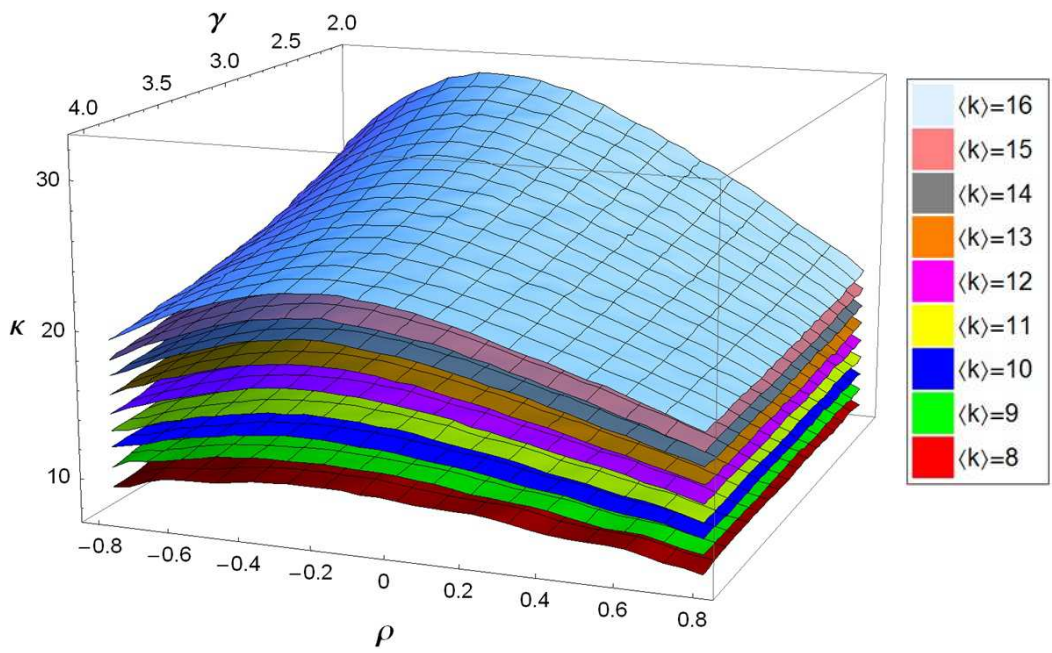}}
\caption{Optimal $\kappa$ degree cutoff values (that minimize the size of CDS) as a function of degree exponent $\gamma$ and Spearman's $\rho$. Each layer represents different average degrees, in uncorrelated (cCONF) networks with $N=2000$. Data is averaged over $200$ network realizations. Data grid resolution: $\Delta \gamma=0.05$, $\Delta \rho=0.05$.}
\label{fig-3D}
\end{figure}

We also analyze the effects of assortativity on the \textit{optimal} $\kappa$ degree threshold value that minimizes the size of CDS. Figure \ref{fig-3D} provides a complete dependence map of the optimal $\kappa$ with respect to two vital network parameters: power-law degree exponent $\gamma$, and assortativity, measured by Spearman's $\rho$. Regardless of $\gamma$ and $\rho$, we can see that $\kappa$ is roughly proportional to the network's average degree. Also, we observe that for any particular network assortativity (and $\rho$ value), $\kappa \sim e^{-\gamma}$. However, it is intriguing that for a fixed $\gamma$ value, $\kappa$ has a maximum approximately at $\rho=-0.4$.

\section*{Discussion}

Our first results revealed that the numerically computed size of RDS with uniform node selection probability is much smaller than the upper bound provided by Alon and Spencer \cite{Alon_2000}. Since their bound assumes that all nodes not dominated by the $X$ set are, in the worst case, nodes of the smallest degree, the difference between their bound and our result shows the relative number of nodes that have higher degree neighbors, yet not dominated. In scale-free networks, we indeed expect to find a significant number of lowest degree nodes with high-degree neighbors (especially in disassortative networks), explaining our observations.

It is also remarkable that RDS with optimally chosen $p$ parameter can always provide a smaller dominating set than a simple degree-ranked node selection. While the latter may be favored for its simplicity and plausibility to be effective in heterogeneous networks, our results show that it is not the case; the usefulness of degree-ranked dominating sets beyond theoretical studies is very limited.

The cutoff dominating set (CDS), proposed as a limiting case of RDS with degree-dependent node selection probability, is proven to be a very effective dominating set selection method. Given full network information, a sequential implementation of the algorithm finds CDS for all possible $\kappa$ degree threshold values in $O(E)$ time. However, since the algorithm only uses local connectivity information, a distributed version can be easily designed, for large networks. Further, the value of optimal $\kappa$ (that minimizes the CDS size) has little dependence on particular network parameters, as shown in Fig.~\ref{fig-3D}, thus it can be estimated easily if detailed network information is not available. Based on our extensive numerical simulations, we conjecture that using the optimal $\kappa$ the CDS size is the smallest of all degree-dependent RDS (with any $\beta$), and it approaches the MDS size provided by the greedy algorithm, irrespective of the network's distinct topological properties. This conjecture is further validated by Fig.~\ref{fig-real} that presents results on several real-world network samples.

We can also understand CDS as a method that bridges the degree-ranked and greedy dominator selection methods. When selecting the very first nodes of the dominating sets, both greedy and degree-ranked methods start by selecting the highest degree nodes. Later, they diverge; the degree-ranked selection continues with the high-degree nodes, while greedy specifically seeks out nodes that increase domination maximally, typically smaller degree nodes. The degree-ranked selection eventually becomes very inefficient only because of the presence of low degree nodes connected only to each other (and hard to reach). Thus, degree-ranked selection is efficient at first, but there is a point at which the method should abandon such selection and instead look
for nodes that are still not dominated, and target them specifically. This is exactly what CDS does: it is essentially a degree-ranked selection until $\kappa$ is reached (set $X$), and then the remaining undominated part is simply added as dominators (set $Y$).


While the analytical estimates for RDS and CDS are highly accurate, they are only applicable to uncorrelated scale-free networks. However, the base formula (Eq.~\ref{base}) can be used for any network (not only scale-free), if the degree distribution and degree correlations can be expressed (or approximated) by some formula. Without analytical expressions, one can still calculate the base formula numerically, using observed (sampled) estimates of the degree distribution and degree correlations, assuming that collecting these esimates requires less time than actually running the RDS or CDS algorithms, or if full adjacency information is not available.

The accuracy of our analytical estimates for RDS and CDS seem to be lower for low $\langle k \rangle$ and $\gamma$ values. This inaccuracy is an artifact of our average degree control method, which controls $\langle k \rangle$ by adjusting $k_{\min}$, and removing a certain fraction of smallest degree nodes. The latter becomes significant when $k_{\min} \rightarrow 1$ (for low $\langle k \rangle$), because it causes a slight deviation from a perfect power-law degree distribution. In order to use the analytical formulas (which are very sensitive to $k_{\min}$), we have to estimate a fractional $k_{\min}$, as if it were a cutoff of a continuous and otherwise perfectly satisfied power-law distribution. In reality, we deviate  from power-law, leading to the inaccurate estimates. However, as $\langle k \rangle$ increases, $k_{\min}$ also increases, and the relative deviation from a perfect power-law decreases, hence the increased accuracy. The implication for real networks is that we can expect similarly less accurate estimates if the degree distribution deviates from power-law.


Our numerical study of dominating set sizes with respect to assortativity reveals a general tendency that the dominating set becomes larger as assortativity increases. We can understand this easily. In case of a disassortative network, high degree nodes connect mostly to low degree nodes, therefore we can expect small dominating sets, due to efficient domination via high-degree nodes. In fact, when $\gamma < 2$ scale-free networks may become so disassortative that star subgraphs form and the size of MDS becomes $O(1)$ \cite{Molnar_2013}. On the other hand, hubs are less effective in dominating assortative networks, since most of their connections are used to connect to other high degree nodes. Therefore, the impact of assortativity on each dominating set selection method depends on how much the method relies on high-degree nodes as dominators. This is why the degree-ranked selection shows the worst performance on highly assortative networks, followed by the degree-dependent RDS (and its limiting case, the CDS), which also favors high-degree nodes. Since technological scale-free networks tend to be disassortative, and although social networks tend to be assortative, extreme assortativity is rare, we can safely conclude that CDS is a viable alternative of greedy selection for most scale-free networks.


In summary, we explored probabilistic dominating set selection strategies in scale-free networks with respect to various network properties. We found that as a particular limiting case of degree-dependent random node selection, a deterministic cutoff dominating set (CDS) provides the smallest dominating set among probabilistic methods, and is widely applicable to heterogeneous networks. Even if full adjacency information is not available, the size of CDS (and RDS) can be accurately predicted using our analytical estimates.

\section*{Methods}

We construct our ensembles of synthetic scale-free networks (undirected and unweighted) using the configuration model \cite{Molloy,Britton}. First, we generate a power-law degree distribution with the desired power-law exponent and average degree. The latter is controlled by adjusting the minimum degree cutoff of the distribution, while we always keep the maximum degree cutoff $k_{\max}$ fixed: either $k_{\max}=N-1$ (the maximum possible in any network, hence essentially unrestricted) or $k_{\max}=\sqrt{N}$ (structural cutoff, making the network uncorrelated \cite{Catanzaro,Boguna}). We obtain a degree sequence from the degree distribution by inverse transform sampling. Given the degree sequence, the configuration model assigns the corresponding number of half-edges (stubs) to each node, and connects randomly (uniformly) chosen pairs of stubs to form links between nodes. This procedure is repeated until there are no free stubs left. The result is a multigraph; however, we convert multiple links to single links and remove self-loops to obtain a simple graph. Due to this pruning of multiple links we have some loss of edges, but since we generate networks with $\gamma>2$, this loss is negligible. We have used the same network construction method in our previous work \cite{Molnar_2013} (including the method of controlling the average degree by selecting the proper $k_{\min}$ value from a precomputed lookup table); according to our previous notation we have here CONF and cCONF networks.

We use two types of dominating sets for comparison with probabilistic dominating set selection methods. The first one is an approximation of the MDS, found by a sequential greedy algorithm. This method selects nodes one by one, at each step selecting a node that provides the maximal increase in the number of dominated nodes in the network (with random tie-breaking); this is the same method as used in \cite{Molnar_2013}. The second method is the degree-ranked dominating set selection (DDS), where we build the dominating set by selecting  nodes in decreasing order of degree (with random tie-breaking) until the selected set dominates the entire network.

To find a probabilistic dominating set (with any particular node selection probability), we use the following algorithm. First, we initialize set $Y$ to contain all nodes of the network, and initialize set $X$ to an empty set. Then, we visit each node exactly once, and determine whether it should be added to set $X$, based on the current node selection probability. If so, then we add the current node to set $X$, remove it from set $Y$ (if present), and also remove all of its neighbors present in set $Y$. This way, once all nodes have been evaluated, we obtain the probabilistic dominating set as $X \cup Y$. We use hashed sets for $X$ and $Y$, which makes the addition, check of containment, and removal of nodes from the sets an $O(1)$ time operation (in amortized time). We loop over all nodes exactly once, and visit all their neihbors, therefore we visit each edge exactly twice, making the algorithm's time complexity $O(E)$ and memory complexity $O(N)$, where $E$ is the number of edges and $N$ is the number of nodes in the network. Note, for sparse networks with small average degree, $O(E) = O(N)$.

When we calculate a cutoff dominating set (CDS), there is an additional optimization we use to find the CDS size for \textit{all} possible $\kappa$ degree cutoff values, including the optimal one that minimizes CDS size, in the same time complexity as finding CDS for only one $\kappa$ value. First, we sort nodes into degree classes in $O(N)$ time using counting sort (or bucket-sort). The linear time complexity comes from the fact that both the number of nodes and the range of their degree values is $O(N)$. Then, we loop over all degree classes in decreasing order of degree, and for each degree class we add all nodes to set $X$ (and remove them and their neighbors from set $Y$ at the same time). This way, we can check the value of $|X|+|Y|$ after finishing each degree class, which is exactly the size of CDS with $\kappa$ equal to the current class degree. We can either output the size of CDS at the current degree, or simply record which CDS size at which $\kappa$ was the smallest. Since we process each node exactly the same way as in RDS (except for the specific order in which they are processed), we have the same $O(E)$ time complexity, and it is not increased by the $O(N)$ time needed to sort the nodes.

We control assortativity by randomly mixing the network's edges, using a Markov-chain of double-edge swaps with biased acceptance probabilities. Without the bias, this method was used in \cite{Viger_2005} and \cite{Molnar_2013} to sample networks with a given degree distribution. Here we use the same method, but we introduce an additional condition for accepting a randomly proposed and otherwise possible edge swap. This acceptance probability is parametrized by a control value $a \in [-1,1]$ that introduces the bias toward accepting a higher or lower fraction of swaps that make the network more assortative, based on its value, in the following way:
\begin{equation}
\Pr(\textrm{accept})=
    \begin{cases}
    a & \mbox{if } a>0 \mbox{ and the swap makes the network more assortative} \\
    -a & \mbox{if } a<0 \mbox{ and the swap makes the network more disassortative} \\
    1-|a| & \mathrm{otherwise,}
    \end{cases}
\label{eq-pr-accept}
\end{equation}
Therefore, we obtain the most disassortative network when $a=-1$ and the most assortative network when $a=1$. However, the relationship between $a$ and any particular assortativity measure, such as Spearman's $\rho$, is non-trivial, as shown in Fig.~\ref{fig-AC}. The detailed description of this method is included in the Supplementary Information, Section S.4.

\section*{Acknowledgments}
We thank T. Nguyen for preparing the network-structure files used in
this research from the Flickr and Foursquare data sets. This work
was supported in part by grant No. FA9550-12-1-0405 from the U.S.
Air Force Office of Scientific Research (AFOSR) and the Defense
Advanced Research Projects Agency (DARPA), by the Defense Threat
Reduction Agency (DTRA) Award No. HDTRA1-09-1-0049, by the National
Science Foundation (NSF) Grant No. DMR-1246958, by the Army Research
Laboratory (ARL) under Cooperative Agreement Number
W911NF-09-2-0053, by the Army Research Office (ARO) grant
W911NF-12-1-0546, and by the Office of Naval Research (ONR) Grant
No. N00014-09-1-0607. The views and conclusions contained in this
document are those of the authors and should not be interpreted as
representing the official policies either expressed or implied of
the Army Research Laboratory or the U.S. Government.

\section*{Author Contributions}
F.M., N.D., \'E.Cz., L.Sz., B.K.S. and G.K. designed the research;
F.M. and N.D. implemented and performed numerical experiments and simulations;
F.M., N.D., \'E.Cz., L.Sz., B.K.S. and G.K. analyzed data and discussed results;
F.M., N.D., \'E.Cz., L.Sz., B.K.S. and G.K. wrote and reviewed the manuscript.

\section*{Additional Information}
Competing financial interests: The authors declare no competing financial interests.


\newpage

\section*{Figures \& Captions}

\nopagebreak

\begin{figure}[h!]
\centerline{\includegraphics[width=\textwidth]{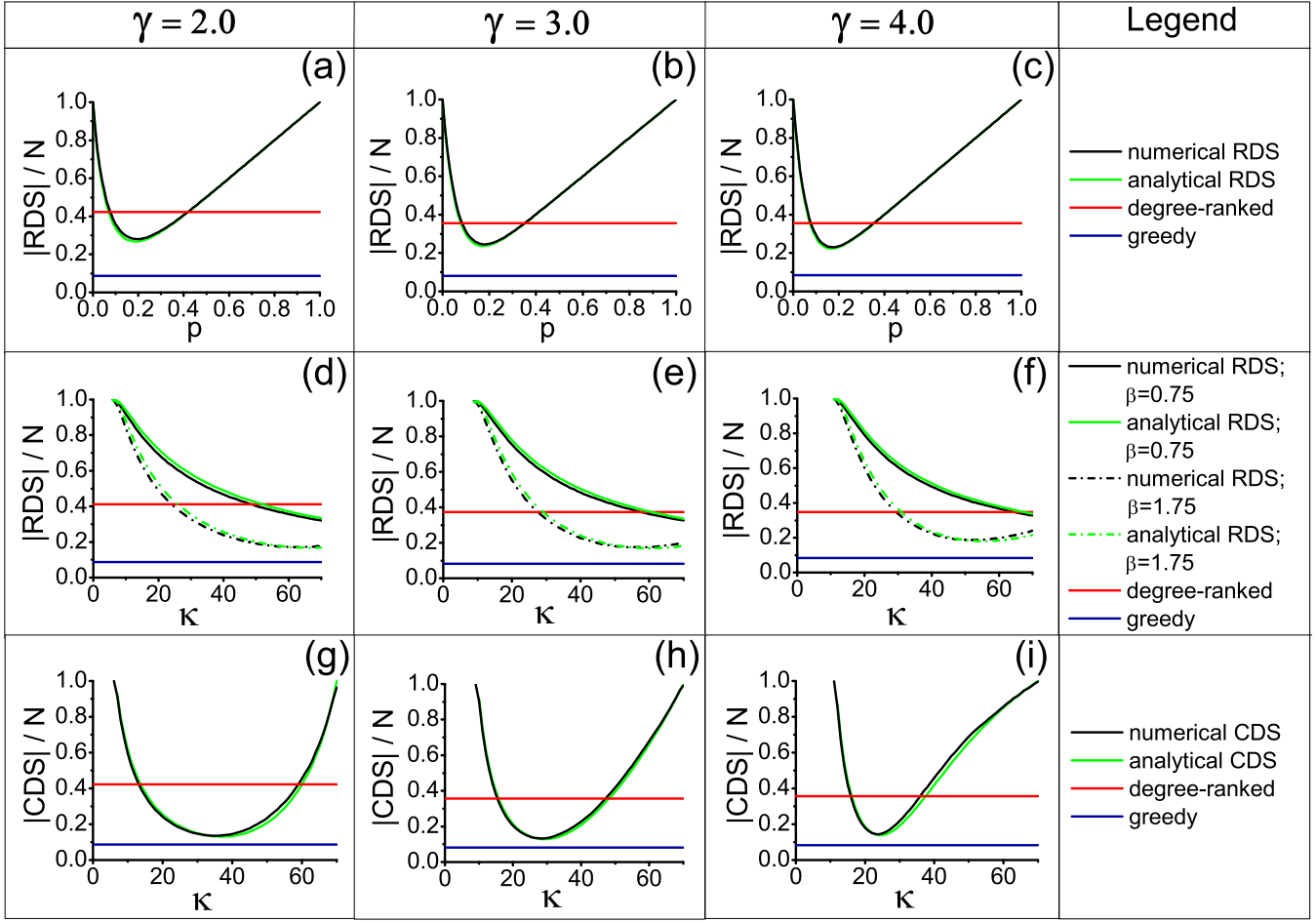}}
\caption{Comparison of analytical estimates [Eqs.~(\ref{RDS}), (\ref{RDS_degree}),and (\ref{CDS})] and numerically computed sizes of RDS and CDS in uncorrelated (cCONF) scale-free networks. For numerical results, data is averaged over 200 network samples. Parameters: $N=5000$ and $\langle k \rangle=16$.}
\label{fig-analytic-real}
\end{figure}

\begin{figure}[h!]
\centerline{\includegraphics[width=\textwidth]{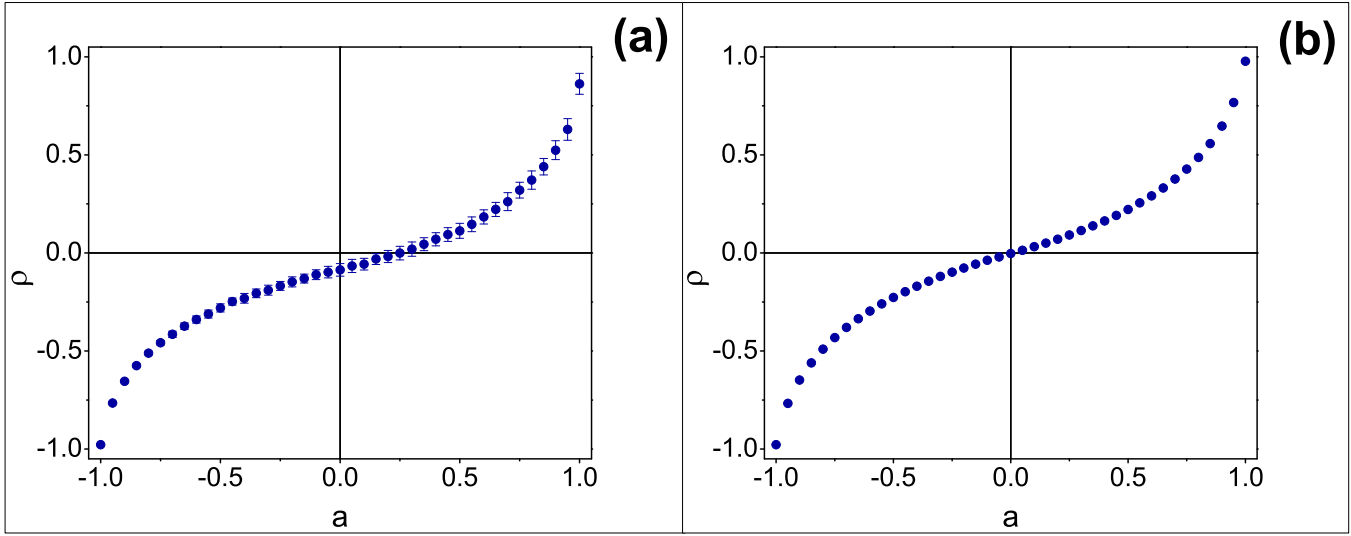}}
\caption{Relationship between the assortativity control parameter $a$ and the achieved Spearman's $\rho$ values. Parameters: $N=5000, \gamma=2.5$, $\langle k\rangle=14$. Data is averaged over $100$ network samples. Error bars indicate the sample standard deviation.}
\label{fig-AC}
\end{figure}

\pagebreak

\end{document}